\documentclass[12pt]{article}
\usepackage{epsfig}
\usepackage{comment}
\usepackage{latexsym}
\usepackage{color}
\newcommand{\mysquare}[0]{\raise-.2ex\hbox{{\Large$\Box$}}}
\def\lsim{\mathrel{\rlap {\raise.5ex\hbox{$ < $}}
{\lower.5ex\hbox{$\sim$}}}}
\def\gsim{\mathrel{\rlap {\raise.5ex\hbox{$ > $}}
{\lower.5ex\hbox{$\sim$}}}} \topmargin -1.5cm \textheight=22.5cm
\catcode`\@=11
\newcount\hour
\newcount\minute
\newtoks\amorpm
\hour=\time\divide\hour by60
\minute=\time{\multiply\hour by60 \global\advance\minute by-\hour}
\edef\standardtime{{\ifnum\hour<12 \global\amorpm={am}%
        \else\global\amorpm={pm}\advance\hour by-12 \fi
        \ifnum\hour=0 \hour=12 \fi
        \number\hour:\ifnum\minute<10 0\fi\number\minute\the\amorpm}}
\edef\militarytime{\number\hour:\ifnum\minute<10 0\fi\number\minute}
\def\draftlabel#1{{\@bsphack\if@filesw {\let\thepage\relax
   \xdef\@gtempa{\write\@auxout{\string
      \newlabel{#1}{{\@currentlabel}{\thepage}}}}}\@gtempa
   \if@nobreak \ifvmode\nobreak\fi\fi\fi\@esphack}
        \gdef\@eqnlabel{#1}}
\def\@eqnlabel{}
\def\@vacuum{}
\def\draftmarginnote#1{\marginpar{\raggedright\scriptsize\tt#1}}
\def\draft{\oddsidemargin -.2truein
        \def\@oddfoot{\sl preliminary draft \hfil
        \rm\thepage\hfil\sl\today\quad\militarytime}
        \let\@evenfoot\@oddfoot \overfullrule 3pt
        \let\label=\draftlabel
        \let\marginnote=\draftmarginnote
   \def\@eqnnum{(\theequation)\rlap{\kern\marginparsep\tt\@eqnlabel}%
\global\let\@eqnlabel\@vacuum}  }

\newcommand{\ba}[0]{\begin{eqnarray}}
\newcommand{\ea}[0]{\end{eqnarray}}
\def\bs{\begin{subequations}}
\def\es{\end{subequations}}

\def\thebibliography#1{%
\vskip 0.5cm \centerline{\bf References}
\list{%
[\arabic{enumi}]}{\settowidth\labelwidth{[#1]}
\leftmargin\labelwidth
\advance\leftmargin\labelsep
\usecounter{enumi}}
\def\newblock{\hskip .11em plus .33em minus .07em}
\sloppy\clubpenalty4000\widowpenalty4000
\sfcode`\.=1000\relax}

\renewcommand{\theequation}{\arabic{section}.\arabic{equation}}

\renewcommand{\section}{\setcounter{equation}{0}\@startsection%
{section}{1}{0mm}{-\baselineskip}{0.5\baselineskip}%
{\normalfont\normalsize\bfseries}}

\renewcommand{\subsection}{\@startsection%
{subsection}{2}{0mm}{-\baselineskip}{0.5\baselineskip}%
{\normalfont\normalsize\bfseries}}

\renewcommand{\subsubsection}{\@startsection%
{subsubsection}{3}{0mm}{-\baselineskip}{0.5\baselineskip}%
{\normalfont\normalsize\slshape}}

\def\crbig{\\\noalign{\vspace{3mm}}}

\def\Fint{{\int d^2\theta\,}}

\def\Re{\,{\rm Re}\, }

\def\s{\sigma}

\def\thefootnote{\fnsymbol{footnote}}
\def\es{\end{subequations}}

\def\ec{\hat E^{c}_{2}}

\newcommand{\uarrw}[0]{\mathrel{
{\raise.5ex\vbox{\hrule width 1cm}\hskip-6pt\rightarrow}}}

\def\bea{\begin{array}}
\def\bem{\begin{displaymath}}
\def\beq{\begin{equation}}

\def\eea{\end{array}}
\def\eem{\end{displaymath}}
\def\eeq{\end{equation}}

\def\ov{\overline}

\def\Re{\mathop{\, \rm Re}\, }

\def\s2w{\sin^2 \theta_W}
\def\Tr{\mathop{\rm Tr}}

\def\crbig{\\\noalign{\vspace {3mm}}}

\def\be{\begin{equation}}
\def\ee{\end{equation}}
\def\bc{\begin{center}}
\def\ec{\end{center}}
\def\bea{\begin{eqnarray}}
\def\eea{\end{eqnarray}}

\begin{document}
\renewcommand{\theequation}{\arabic{section}.\arabic{equation}}

\renewcommand{\section}{\setcounter{equation}{0}\@startsection%
{section}{1}{0mm}{-\baselineskip}{0.5\baselineskip}%
{\normalfont\normalsize\bfseries}}

\renewcommand{\subsection}{\@startsection%
{subsection}{2}{0mm}{-\baselineskip}{0.5\baselineskip}%
{\normalfont\normalsize\slshape}}
\begin{titlepage}
\begin{flushright}
NEIP--05--06 \\
LPTENS--05/19\\
CPTH--RR033.0605\\
hep-th/0601005 \\
December 2005
\end{flushright}

\vspace{8mm}

\begin{centering}
{\bf\LARGE Gaugino Condensates and Fluxes in\\
\vspace{4mm}
{\boldmath{$N = 1$}} Effective Superpotentials$^\ast$}\\

\vspace{9mm}
 {\Large J.-P.~Derendinger,$^1$ C.~Kounnas$^{2}$
and
P.M.~Petropoulos$^3$}

\vspace{6mm}

{\small
$^1$ Physics Institute, Neuch\^atel University, A.-L. Breguet 1,
\\ CH--2000 Neuch\^atel, Switzerland

\vskip .1cm

$^2$ Laboratoire de Physique Th\'eorique,
Ecole Normale Sup\'erieure,$^\dagger$ \\
24 rue Lhomond, F--75231 Paris Cedex 05, France

\vskip .1cm

$^3$ Centre de Physique Th\'eorique, Ecole Polytechnique,$^\diamond$
\\
F--91128 Palaiseau, France  }

\vspace{9mm}

{\bf\Large Abstract}

\end{centering}

\vspace{4mm}

\begin{quote}

In the framework of orbifold compactifications of heterotic and
type II orientifolds, we study effective $N=1$ supergravity
potentials arising from fluxes and gaugino condensates. These
string solutions display a broad phenomenology which we analyze
using the method of $N=4$ supergravity gaugings. We give examples
in type II and heterotic compactifications of combined fluxes and
condensates leading to vacua with naturally small supersymmetry
breaking scale controlled by the condensate, cases where the
supersymmetry breaking scale is specified by the fluxes even in
the presence of a condensate and also examples where fluxes and
condensates conspire to preserve supersymmetry.
\end{quote}

\vspace{5pt}
\vfill
\hrule width 6.7cm \vskip.1mm{\small \small \small \noindent
$^\ast$\ Research partially supported by the EU under the
contracts MEXT-CT-2003-509661,
MRTN-CT-2004-005104 and MRTN-CT-2004-503369.\\
$^\dagger$\ Unit{\'e} mixte  du CNRS et de l'Ecole
Normale Sup{\'e}rieure, UMR 8549.\\
$^\diamond$\  Unit{\'e} mixte  du CNRS et de  l'Ecole
Polytechnique, UMR 7644.}
\end{titlepage}
\newpage
\setcounter{footnote}{0}
\renewcommand{\thefootnote}{\arabic{footnote}}

\setlength{\baselineskip}{.7cm} \setlength{\parskip}{.2cm}

\setcounter{section}{0}
\section{Introduction}

Superstring constructions provide a plethora of four-dimensional
vacua with exact or spontaneously broken supersymmetries. Those
with chiral-fermion families have $N=1$ or $N=0$ supersymmetry in
four dimensions. Such models can be built in the framework of
heterotic compactifications or type II orientifolds with branes and fluxes.

In the case of vacua with $N=1$ supersymmetry, it is expected that
nonperturbative phenomena break the leftover supersymmetry. These
nonperturbative effects are controlled by gaugino condensation
occurring in the infrared regime of strongly coupled gauge sectors
\cite{gaugino1,gaugino2}. Both approaches predict modifications in
the superpotential of the effective supergravity theory. The
leading terms are in general of the form
\begin{equation}
\label{wnp1} W_{\rm nonpert} = \mu^3 \, {\rm exp}\left( - {24\pi^2
Z \over b_0 } \right)  ,
\end{equation}
where $b_0$ is a one-loop beta-function coefficient, $\mu$ a scale
at which the Wilson coupling $g^2(\mu)$ is defined and $Z$ a
modulus such that $\Re Z = g^{-2}(\mu)$. The expectation value of
the nonperturbative superpotential defines the
renormalization-group-invariant transmutation scale $\Lambda$ of
the confining gauge sector in which gauginos condense, $\langle
W_{\rm nonpert.} \rangle= \Lambda^3$. In a given string
compactification, physical quantities are functions of moduli
fields. In particular, {\it the scale $\mu$ itself is in general a
modulus-dependent quantity}. Typically, the exponent in the
nonperturbative superpotential (\ref{wnp1}) is a number of order
ten or more and $N$-instanton corrections ($N>1$) are
exponentially suppressed. But several condensates could form and
the nonperturbative superpotential could include several similar
terms involving various moduli.

Which modulus $Z$ appears in Eq.~(\ref{wnp1}) depends on the
underlying string (or M-) theory. In the heterotic string, it is
identified with the dilaton field $S$ \cite{gaugino2, het}. In
type II orientifolds, $Z$ is a well-defined combination of $U$ and
$S$ in IIA theories and of $T$ and $S$ in IIB (or F-theory)
compactifications \cite{LRS, DKPZ, DKPZlong}.

These nonperturbative contributions coexist, in the effective
superpotential, with the perturbative moduli-dependent terms
produced by compactifications of heterotic \cite{gaugino2,
hetflux}, type IIA \cite{IIAflux} and type IIB \cite{IIBflux}
compactifications with various types of fluxes. Recently, a method
has been developed to determine unambiguously the K\"ahler
potential and the superpotential of the effective supergravity in
terms of \emph{all} fluxes present in the fundamental theory
\cite{DKPZ}. This allows for an exhaustive analysis, in particular
for the type IIA where no systematic Calabi--Yau approach exists
due to the presence of torsion induced by geometrical fluxes ({\it
``generalized" Calabi--Yau}) \cite{CYgen}. This includes a large
class of (freely acting) orbifolds or fermionic string models
\cite{ferm}. The principle of the method is that the $N=1$ theory
at hand is strongly constrained by the universality properties of
the underlying $N\geq 4$ extended supergravity, inherited from the
string fundamental theory before the orbifold projection. The only
available tool for generating a non-trivial potential in $N\geq 4$
supergravity theories is the \emph{gauging procedure} \cite{N=4},
ultimately related to the nature and strength of the fluxes
present in the theory. The gauging procedure determines completely
the $N=1$ superpotential, at least for the bulk fields. Those are
relevant for the determination of the vacuum structure, before
considering the usual perturbative corrections in the observable
sector, originating from the renormalization of the softly broken
$N=1$ supersymmetry.

Although the inclusion of the nonperturbative corrections in the
flux-induced superpotential has been proposed by several authors
in early and recent literature \cite{condflux}, the conclusions
have been either controversial or incomplete, mainly due to the
pathological behaviour of the vacuum, like for instance:
\begin{itemize}
\item[(i)] runaway behaviour of the moduli $Z$, \item[(ii)]
destabilization of the no-scale structure/positivity of the
effective potential, \item[(iii)] undesired transitions to anti-de
Sitter vacua, \item[(iv)] fine-tuning problem associated with the
quantization of the flux coefficients.
\end{itemize}
Having the generic and unambiguous structure of the effective
superpotential in the presence of fluxes, we will reexamine these
issues and show that these pathologies of the vacuum can be
avoided with a suitable choice of fluxes in heterotic, IIA and IIB
$N=1$ effective supergravities. As we will see in several
examples, the nonperturbative contribution involving the
superfield $Z$ \emph{is not always} the source for supersymmetry
breaking.

Our paper has mainly two parts. In Sec. \ref{genfra}, we first
recall the generation of $N=1$ superpotentials from fluxes,
following Ref. \cite{DKPZ}. We then summarize the general
framework for describing condensates in an effective Lagrangian
approach and describe some of the common caveats of the method.
Section \ref{ftsgs} is devoted to the study of examples
illustrating several scenarios: (i) situations where supersymmetry
breaking is independent from gaugino condensation and (ii) cases
where the gaugino condensates induce supersymmetry breaking. These
examples fall into three different classes with respect to the
scaling properties of the gravitino mass. The difficulties one
usually encounters when including the nonperturbative corrections
are clearly described. The resolution of these problems is
possible provided one realizes that the issues of moduli
stabilization, supersymmetry breaking, gaugino condensation and
positivity of the potential, although related, must be treated
\emph{separately}. Numerous examples can be found both in type II
and heterotic where supersymmetry breaking is driven by fluxes,
the condensate playing a subdominant role. One can also exhibit
generic type II models where the gaugino condensate itself is
responsible for the supersymmetry breaking. Similar realizations
exist in heterotic string provided the fluxes are chosen to be
large, with ratios of order one though. A summary and some
comments are given in the last section.

\section{Superpotentials, gaugings and condensates}\label{genfra}
\subsection{Flux dependence of the superpotential}\label{secflux}

We will confine our discussion to the $Z_2\times Z_2$ orbifold
compactification of heterotic strings, or IIA and IIB
orientifolds. This compactification setup leads to seven main
moduli (including the string dilaton) and $N=1$ supersymmetry. The
effective supergravity is the $Z_2\times Z_2$ projection of the
$N=4$ theory which would describe the sixteen-supercharge
ten-dimensional theory (on a simple six-torus). Fluxes are
introduced by gauging the $N=4$ supergravity. As shown in Ref.
\cite{DKPZ}, the gauging allows the introduction of NS-NS, R-R
fluxes as well as geometric fluxes \cite{DKPZ, ALT} corresponding
to non-trivial internal spin connections.

Let us restrict ourselves to the main moduli, namely the string
coupling, the six geometrical moduli fields and their
superpartners. Their precise spelling depends on the theory under
consideration. For heterotic strings on ${T^6/ Z^2\times Z^2}$,
these are the dilaton--axion superfield $S$, the volume moduli $T_A$ and the
complex-structure moduli $U_A$, $A=1,2,3$. The index $A$ refers to the three
complex planes left invariant by $Z_2\times Z_2$.
The $N = 1$ supersymmetric complexification for these fields is defined naturally
in terms of the geometrical moduli $G_{ij}$ (nine fields), the string dilaton $\Phi$
and the components $B_{ij}$ (three fields) and $B_{\mu\nu}\sim a$ of the
antisymmetric tensor. Explicitly, the metric tensor restricted to the plane $A$ is
\beq
\label{moduli1}
\left( G_{ij} \right)_A =\frac{t_A}{u_A}
\left(\begin{array}{cc}
u_A^2 +\nu_A^2 & \nu_A \\
\nu_A  & 1 \
\end{array}\right) ,
\eeq with \be \label{moduli2} T_A=t_A+i\left( B_{ij}\right)_A,
\qquad\qquad  U_A= u_A + i \, \nu_A \ee and \be \label{moduli3}
{\rm e}^{-2\Phi} = s(t_1t_2t_3)^{-1},  \qquad\qquad S=s+ ia. \ee
The Weyl rescaling to the four-dimensional Einstein frame is
$G_{ij}=s^{-1} {\tilde G}_{ij}$. The $Z_2\times Z_2$ projection of
the $N=4$ theory leads to the scalar K\"ahler manifold
\begin{equation}
M=\left[ { SU(1,1) \over U (1)} \right]_S \times \prod_{A=1}^3
\left[ { SU(1,1) \over U (1)} \right]_{T_A} \times \prod_{A=1}^3
\left[ {SU(1,1) \over U (1)} \right]_{U_A} ,
\end{equation}
with K\"ahler potential (in the usual string parameterization)
\begin{equation}
\label{Kis} K = - \log \left(S+{\ov S}\right) -\sum_{A=1}^3 \log
\left(T_A+{\ov T}_A\right)\left(U_A+{\ov U}_A\right).
\end{equation}
Coupling the seven moduli superfields to further multiplets $Z_A^I$,
including gauge-charged states, modifies however the K\"ahler potential of the
$N=1$ theory, to become
\begin{equation}
\label{Kis2} K = - \log \left(S+{\ov S}\right) -\sum_{A=1}^3 \log
\left[\left(T_A+{\ov T}_A\right)\left(U_A+{\ov U}_A\right) -
\sum_{I=1}^{n_A} \left(Z_A^I+ \ov Z_A^I\right)^2 \right].
\end{equation}

The superpotential, as explained in Ref. \cite{DKPZ}, is
determined from the gravitino mass term of the $N=4$ gauged
supergravity, after the orbifold and orientifold projections: \be
\label{superpot}
 {\rm e}^{K/2} \, W= {1\over6} \epsilon^{ABC} \left[
 (\phi_0 -\phi_1 ) \, f_{IJK}\,  +(\phi_0
 +\phi_1)\, {\ov f}_{IJK} \right] \, \Phi^I_A\,  \Phi^J_B\, \Phi^K_C,
 \ee
where the antisymmetric numbers $f_{IJK}$ and $\ov f_{IJK}$ are
the structure constants of the $N=4$ gauged algebra multiplied by
the $SO(6,n)$ metric and $SU(1,1)$ $S$-duality phases
\cite{N=4}\footnote{The indices $I,J,K$ label the vector
representation of the $SO(6,n)$ global symmetry of the (ungauged)
$N=4$ theory and $\eta_{IJ}$ is the invariant metric for this
representation. Then, for instance, $f_{IJK} = {f_{IJ}}^L
\eta_{LK}$, where ${f_{IJ}}^L$ is a structure constant.}. The
constrained fields
 \be
 \phi_0, \, \phi_1 \quad {\rm and} \quad \Phi^I_A=\left\{
 \sigma^1_A,\sigma^2_A ; \rho^1_A,\rho^2_A, \chi_A^I \right\}
\ee
define the scalar manifold as the solution of the following constraint equations:
\beq \label{Poin1}
\begin{array}{rcl}
{1\over 2} &=& \left|\phi_0\right|^2 - \left|\phi_1\right|^2 ,
\crbig { 1\over 2} &=& \left|\sigma_A^1\right|^2 +
\left|\sigma_A^2\right|^2 - \left|\rho_A^1\right|^2 -
\left|\rho_A^2\right|^2 - \sum_I \left|\chi_A^I\right|^2 \, ,
\crbig 0 &=& \left(\sigma_A^1\right)^2 + \left(\sigma_A^2\right)^2
- \left(\rho_A^1\right)^2 - \left(\rho_A^2\right)^2 - \sum_I
\left(\chi_A^I\right)^2 \,, \quad\qquad (A=1,2,3)\,.
\end{array}
\eeq

Since our goal is to work with a fixed set of well-defined moduli
fields $(S,T_A, U_A)$, and to study various classes of gaugings
induced by the fluxes, it is necessary to solve the above
constraints in terms of the fields $(S,T_A, U_A, Z_A^I)$:
\begin{itemize}
\item $S$ manifold  $\left[ SU(1,1) / U (1) \right]_S$:
\be \phi_0-\phi_1 ={1\over \left(S+\ov S\right)^{1/2}}, \qquad
\phi_0+\phi_1 ={S\over \left(S+\ov S\right)^{1/2}}; \ee
\item $T_A, U_A , Z_A^I$ manifold
$\left[  SO(2,2+ n_A ) / SO(2) \times SO(2+n_A ) \right]_{T_A,U_A,Z_A^I}$:
\beq \label{solution}
\begin{array}{rclrcl}
\sigma_A^1 &=& {1\over 2Y_A^{1/2}} \, \left( 1 + T_AU_A -
\left(Z_A^I\right)^2 \right) \,, \qquad&\qquad \sigma_A^2 &=&
{i\over 2Y_A^{1/2}} \, \left( T_A + U_A \right) \,, \crbig
\rho_A^1 &=& {1\over 2Y_A^{1/2}} \, \left( 1 - T_AU_A +
\left(Z_A^I\right)^2 \right) \,, & \rho_A^2 &=& {i\over
2Y_A^{1/2}} \, \left( T_A - U_A \right) \,, \crbig \chi_A^I &=&
{i\over Y_A^{1/2}} \, Z_A^I \,,
\end{array}
\eeq where $Y_A = \left(T_A+\ov T_A\right)\left(U_A + \ov
U_A\right) - \sum_I\left(Z_A^i + \ov Z_A^I\right)^2$.

\end{itemize}
The above equations allow to rewrite the scalar potential and the
gravitino mass term as functions of the $N=1$ complex scalars, the
perturbative $f_{IJK}$-term as well as the nonperturbative
$S$-dual term, $S\, {\ov f}_{IJK}$. The K\"ahler potential and the
superpotential can then be obtained by separating the holomorphic
part in the $N=1$ gravitino mass term, using the relation $m_{3/2}
= {\rm e}^{K/2} W$. The result is the K\"ahler potential given
above in Eq.~(\ref{Kis2}).

It is important to realize that even if the scalar manifolds for
type IIA or IIB orientifold compactifications are the same as for
heterotic strings, the identification of the complex scalar fields
present in the seven chiral multiplets $S$, $T_A$, $U_A$ changes.
This is obvious for the imaginary parts since tensor fields change
with the ten-dimensional string theory. For the real parts, the
relation between the seven main moduli $s$, $t_A$ and $u_A$, as
defined from the metric and the dilaton in
Eqs.~(\ref{moduli1}--\ref{moduli3}), and the real parts of $S$, $T_A$
and $U_A$ is a (non-supersymmetric) field redefinition which
leaves invariant the kinetic terms\footnote{The kinetic terms for
the main moduli $t_A$ and $u_A$ arise from the compactification of
the Einstein term and are thus universal.}. The appropriate field
redefinition is dictated by the dilaton rescaling and the
orientifold projection. Consider for instance the type IIA
orientifold with $D_6$-branes. While the superfields $T_A$ are the
same as for heterotic strings, $S$ and $U_A$ are now defined by
\beq \label{IIA moduli1}
\begin{array}{c}
S = s' + i A_{6810}, \crbig U_1 = u_1'- iA_{679}, \quad U_2 = u_2'
- iA_{589},  \quad U_3 = u_3' - iA_{5710},
\end{array}
\eeq in terms of the $(Z_2\times Z_2)$-invariant components of the
R-R three-form $A_{ijk}$ ($i,j,k= 5,\ldots,10$). The relations
\beq \label{chIIA}
 s'=\sqrt{s\over u_1u_2u_3},\qquad
 u_1'=\sqrt{su_2u_3\over u_1}, \qquad u_2'=\sqrt{su_1u_3\over
 u_2},\qquad u_3'=\sqrt{su_1u_2\over u_3}
\eeq define then the real parts in terms of the geometric moduli
introduced in Eqs.~(\ref{moduli1}--\ref{moduli3}).

We can illustrate the above in the $T^6/Z_2$ type IIB orientifold
with a $Z_2\times Z_2$ orbifold projection. The $Z_2$ inverts all
six internal coordinates and this is compatible with the presence
of $D_3$-branes. A field redefinition similar to
Eqs.~(\ref{chIIA}) but involving $s$ and $t_A$ is again necessary. The
NS-NS and R-R three-forms generate fluxes compatible with the
$Z_2\times Z_2$ setup. These fluxes lead to the following
superpotential terms: \beq \label{IIB}
\begin{array}{lll}
\makebox{R-R three-form:} & 1 &\makebox{(from $F_{579}$), }
\crbig & i(U_1+U_2+U_3) &\makebox{(from
$F_{679}=F_{589}=F_{5710}$), } \crbig & -(U_1U_2+U_2U_3+U_3U_1)
&\makebox{(from $F_{689}=F_{5810}=F_{6710}$), } \crbig &
-iU_1U_2U_3 &\makebox{(from $F_{6810}$),} \crbig \makebox{NS-NS
three-form:} & iS &\makebox{(from $H_{579}$), } \crbig &  -
S(U_1+U_2+U_3) &\makebox{(from $H_{679}=H_{589}=H_{5710}$), }
\crbig & -iS (U_1U_2+U_2U_3+U_3U_1) &\makebox{(from
$H_{689}=H_{5810}=H_{6710}$), } \crbig & SU_1U_2U_3
 &\makebox{(from $H_{6810}$).}
 \end{array}
 \eeq
More fluxes are allowed in type IIA or heterotic models, where
also geometric fluxes\footnote{Equivalent to background values of
spin-connection components.} are permitted by the orbifold
projection. These fluxes would induce other (possibly
$T_A$-dependent) terms in the superpotential. Those are classified
in \cite{DKPZ}, and we will use these terms in the examples
presented in Sec. \ref{ftsgs}.

\subsection{Effective field theory description of condensates }\label{seccond}

The generation of a nonperturbative superpotential due to gaugino condensates
follows schematically from the following argument.
In $N=1$ supersymmetric field theories,
the coupling of gauge fields to chiral multiplets involves a holomorphic function,
$$
{1\over4} \Fint f\left(\phi^i, M^a\right) WW + {\rm h.c.},
$$
where $\phi^i$ denotes the charged matter chiral superfields and $M^a$ the (gauge-neutral)
moduli fields.
In the Wilson Lagrangian derived from a superstring theory, the function $f$ receives
contributions from several sources, including renormalization-group (one-loop) running
and field-dependent threshold corrections due to charged matter and moduli fields.

To obtain the effective action for gaugino condensates, one first
introduces a composite classical chiral superfield $U \simeq
\langle WW \rangle$. The dynamics of $U$ is then (mostly) dictated
by anomaly-matching. Gauge kinetic terms are replaced by an
$F$-density Lagrangian with in general three contributions:
\beq
\label{Fdens} {1\over4}\Fint \left[ \, f_{\rm eff}
\left(M^a,\phi^i\right) \, U + w_{VY}(U) + w_{K}\left(M^a, \phi^i\right)
\, U \, \right] + {\rm h.c.}
\eeq
Suppose that the (string
tree-level) gauge coupling field is related to the modulus $S$,
$\Re S = g_W^{-2}$. The holomorphic function $f_{\rm eff}$ is
then\footnote{Up to irrelevant constants if the gauge group is not
simple.}
\beq \label{feffis} Z\equiv f_{\rm eff} = S + f_{\rm
thr}\left(M^a , \phi^i\right),
\eeq
where $f_{\rm thr}$ would
describe moduli-dependent threshold corrections. The modulus $Z$
appearing in the introduction is actually $f_{\rm eff}$. The
leading behaviour of typical string theory threshold corrections
is linear in some combination of moduli fields \cite{threshold},
or logarithmic \cite{Mariosthreshold}. Field theory thresholds,
which in general have a logarithmic behaviour, are actually
described by $w_K$, as discussed below. The second contribution in
expression (\ref{Fdens}) is the Veneziano--Yankielowicz
superpotential matching the anomaly of the superconformal chiral
$U(1)$ \cite{VY}: \beq \label{VYis} w_{VY}(U) = {b_0\over 24\pi^2}
U\left[ \ln \left({U\over\mu^3}\right) - 1 \right], \eeq where
$b_0 = 3C(G)-T(R)$ is the one-loop beta-function coefficient and
$\mu$ is the scale at which the Wilson action has been
defined\footnote{The scale $\mu$ is the ultra-violet cutoff of the
wilsonnian action.}, $\Re S  = g_W^{-2}(\mu)$. Finally, Konishi
anomalies \cite{Konishi} related to chiral $U(1)$ (non $R$-)
symmetries acting on each charged matter representation generate
the last term, which more precisely reads
$$
{1\over4} \sum_I {b_I\over24\pi^2}  \Fint U \ln \left( {{\cal P}_I(\phi^i, M^a)
\over {U}^{p_I} }  \right) + {\rm h.c.}
$$
It does not depend on the Wilson scale $\mu$\footnote{It is an
``infrared" contribution describing in the wilsonnian action the
physics of spontaneous symmetry breakings induced by chiral
fields.}. The sum is over the irreducible representations of the
chiral superfields and $b_I$ is a positive coefficient dictated by
anomaly-matching. This term can also be written as
$$
{1\over96\pi^2}  \Fint U \ln \left( {{\cal P}(\phi^i, M^a) \over {U}{^p_I} } \right)
+ {\rm h.c.},
\qquad\qquad
{\cal P}(\phi^i, M^a)  = \prod_I {\cal P}_I(\phi^i, M^a)^{b_I},
$$
with $p = \sum_I b_Ip_I$. The quantity ${\cal P}$ is an analytic
gauge invariant polynomial which characterizes matter condensates
along the $D$-flat directions of the scalar potential and the
exponent $p$ is such that the argument of the logarithm is
dimensionless \cite{BDFS, DZ}.

The nonperturbative superpotential is then obtained by eliminating
$U$ in the $F$-density, with the result (if $p \ne b_0$) \beq
\label{Wnonpert} W_{\rm nonpert} = W_{\rm
classical}\left(M^a,\phi^i\right) + \mu^3 \, {\rm e}^{{p\over
b_0-p}} \,\left({{\cal P}\over \mu^{3p}}\right)^{{1\over p-b_0}}
\, {\rm exp}\,\left[ - {24\pi^2 \over b_0-p} ( S + f_{\rm thr})
\right]. \eeq For a symmetry breaking $G \rightarrow  H$, the
positive number $p$ is given by $\left. b_0 \right|_G - \left. b_0
\right|_H = 2T(R_{\rm Goldstone})$, where $R_{\rm Goldstone}$ is
the representation of the coset $G/H$. Notice that the term
matching Konishi anomalies modifies the Veneziano--Yankielowicz
term. This modification is precisely required by the threshold
induced when chiral fields break gauge symmetries\footnote{These
are field theory thresholds induced in the range of validity of
the Wilson Lagrangian, at scales lower than the ultra-violet
cutoff $\mu$.}.

The terms matching Konishi anomalies and inducing field theory
thresholds produce then an effective, field-dependent scale \beq
\label{mueff} \mu^3_{\rm eff} \left(M^a,\phi^i\right) = \mu^3 \,
{\rm e}^{{p\over b_0-p}} \,\left({{\cal P}\over
\mu^{3p}}\right)^{{1\over p-b_0}} \eeq in front of the usual
superpotential with exponential behaviour in the gauge coupling
field $S$ supplemented by string threshold contributions.

The particular case $p=b_0$ would eliminate completely the $U\ln
U$ contribution in the effective superpotential. As an example,
consider $N=2$ super-Yang--Mills theory. In terms of $N=1$
superfields, the theory includes a single adjoint chiral
superfield which generically breaks the gauge group to its abelian
Cartan subgroup. Using then ${\cal P}_I(\phi) = \Tr(\phi^2)$,
anomaly matching requires $p = 2C(G) = b_0$ (and $b_I=3/2$). As a
result, the effective superpotential is simply
$$
{1\over4} \, U \, \left[ S + {b_0\over16\pi^2} \ln\left( {\Tr (\phi^2) \over \mu^2}
\right) \right]
$$
and eliminating $U$ cancels the nonperturbative superpotential
with however \beq \label{N=21} \Tr(\phi^2) = \mu^2 {\rm e}^{ -
16\pi^2 S / b_0}. \eeq In a general $N=1$ case with $p=b_0$, \beq
\label{p=b0case} W_{\rm nonpert} = W_{\rm classical}(M^a,\phi^i)
\eeq and the scale of the invariant ${\cal P}$ is fixed by the
equation \beq \label{p=b0P} {\cal P} (\phi^i, M^a) = \mu^{3b_0}
{\rm e}^{-24\pi^2\left[S + f_{\rm
thr}\left(M^a,\phi^i\right)\right]} . \eeq Even if the condensate
contribution to the nonperturbative superpotential is cancelled,
the matter fields show an exponential behaviour in the gauge
coupling field. This behaviour will explicitly reappear in the
coupling, via the classical superpotential, of matter fields
$\phi^i$ to gauge singlet fields (like moduli fields) or to
``spectator" charged matter superfields. Again in the $N=2$
example, hypermultiplets couple to the gauge multiplet via the
(classical) superpotential term $\sqrt2 H^c \phi H$. Expression
(\ref{N=21}) generates then a mass term of the form \beq
\label{N=22} \sim \mu {\rm e}^{ - 8\pi^2 S / b_0} H^c T H , \eeq
with some constant matrix $T$.

\subsection{The condensate scale, supersymmetry breaking and the
runaway problem} \label{secproblem}

Let us now review in a simple situation the role of the condensate superpotential
in defining the vacuum structure of the effective supergravity, illustrating the
fine-tuning problem alluded to in the introduction.
Consider the following simplified superpotential:
\begin{equation}
\label{ftW}
W = a + w(S),
\qquad\qquad
w(S)= \mu^3 {\rm e}^{-S}
\end{equation}
and we have for convenience rescaled $S$ according to
$$
{24\pi^2 S \over b_0} \quad\longrightarrow\quad S.
$$
This rescaling leaves the corresponding kinetic terms unchanged
and multiplies the scalar potential by an overall factor. The
quantity $a$ is in general $T_A$- and $U_A$-dependent. It includes
the perturbative contributions to the superpotential induced by
fluxes. In a heterotic compactification with vanishing geometrical
fluxes however, only the $U_A$-dependence generated by the
heterotic three-form $H_3$ does survive. As already stated, the
scale $\mu$ may depend on moduli $U_A$ or $T_A$, but $a$ does not
depend on $S$ in heterotic compactifications.

Due to the $SU(1,1)$ structure of the K\"ahler manifold, the
scalar potential considerably simplifies and takes the following
suggestive form:
\begin{equation}
\label{pot3} {\rm e}^{-K} V = \sum_i \left| W - W_i(Z_i + \ov
Z_i)\right|^2 - 3 |W|^2,
\end{equation}
where $ \{ Z_i \} \equiv \{ S, T_A, U_A \}$ and $W_i =
\partial_{Z_i } W$.

Consider now situations where the geometrical fluxes are indeed
absent and the superpotential is $T_A$-independent. The scalar
potential becomes:
\begin{equation}
\label{nspot} {\rm e}^{-K} V = \sum_{\{ Z_i \} \equiv \{ S, U_A
\}} \left|W - W_i(Z_i + \ov Z_i)\right|^2.
\end{equation}
We are led to a \emph{no-scale model} \cite{noscale}, with a
semi-positive-definite potential and flat directions $\{T_A\}$.
The $U_A$ moduli are generically fixed by their minimization
conditions and $a$ and $\mu^3$ in Eqs.~(\ref{ftW}) are effectively
constant. The remaining minimization condition for the $S$ field,
\begin{equation}
\label{minS} a + \left(S+\ov S + 1\right)w(S)=0,
\end{equation}
determines the value of $S$. Supersymmetry is broken in the
$T_A$-directions, in Minkowski space. Equation (\ref{minS}) shows
that an exponentially small value of $w(S)$ necessarily implies
\beq | a | \ll 1, \eeq a fine-tuning condition on the perturbative
fluxes.

This is a potential problem in cases with $U_A$-independent $a$,
like for instance in the $Z_3$ orbifold where the would-be $U_A$
moduli are frozen to their self-dual values by the orbifold point
group action and $a$ is then given by the (constant) $H_3$-form
flux. Because of flux quantization, the perturbative
superpotential $a$ is therefore of $\mathcal{O}(1)$ or zero and a
non-zero $a$ is in contradiction with the minimum condition. A
vanishing $a$ is also problematic since it leads to a runaway
potential:
\begin{equation}
V = \frac{\left(1 + S + \ov S\right)^2}{\left(S+\ov
S\right)\prod_A\left(T_A + \ov T_A\right)} \, |\mu|^6 \, {\rm
e}^{-\left(S+\ov S\right)} .
\end{equation}
Various attempts have been proposed in the past for improving this
situation (see for instance \cite{VZ}), including multiple gauge group condensations (without
fluxes) which however do not help in removing the fine-tuning
problem with a non-zero $a$.

The above conclusions are drastically modified with a modulus-dependent perturbative
superpotential, as induced by fluxes. For instance, type IIB models with stable moduli
$S$ and $U_A$ have automatically a no-scale structure with all $T_A$ moduli being
flat directions. For example, this can be realized with
\begin{eqnarray}\label{kacpol}
W&=& A\left[ 1 +U_1U_2+U_2U_3+U_3U_1 + \gamma  S(U_1+U_2+U_3 +
U_1U_2U_3)\right] \nonumber \\ &+&i B \left[ U_1+U_2+U_3 +
U_1U_2U_3 + \gamma S(1+ U_1U_2+U_2U_3+U_3U_1) \right],
\end{eqnarray}
where $A, B$ and $\gamma A, \gamma B$ are respectively proportional to
R-R and NS-NS flux numbers. The absence of any $T_A$ dependence in $W$ is
responsible for the appearance of the no-scale structure
with semi-positive-definite scalar potential. This superpotential fixes the moduli
$(U_A, \gamma S)$ to unity and supersymmetry is broken in flat space with
gravitino mass $m_{3/2}^2 \propto (A^2 +B^2)/\prod_A(T_A + \ov T_A)$.
A plethora of examples exists with similar properties, or with
unbroken supersymmetry whenever $W$ vanishes.

We now observe that in this example, one expects nonperturbative
superpotential contributions proportional to $\exp{-\alpha_AT_A}$
arising from condensates that live on $D_7$-branes\footnote{While
contributions proportional to ${\rm e}^{-\alpha S}$ would arise
from $D_3$-branes.}. To be concrete, add to $W$ in
Eq.~(\ref{kacpol}) a nonperturbative contribution $ \exp[- \alpha
(T_1 + T_2 + T_3)] = \exp(- 3 \alpha T)$ (all $T_A$ equal). It is
clear from the above analysis that this addition spoils
\emph{arbitrarily} the no-scale structure and destabilizes the
Minkowski vacuum: the moduli $T$ gets stabilized but the potential
becomes negative, $V = - 3m_{3/2}^2$, as required by unbroken
supersymmetry in anti-de Sitter space. Notice that positivity
would have been lost as well, had $w(T)$ been a function of a
single $T_A$ modulus. In such a case however, the potential does
not have a critical point and the fundamental state of the theory
is described by domain-wall-like solutions with 1/2 supersymmetry.

In this example, the vacuum structure of the theory is clearly only
understood from the analysis of the combined perturbative and nonperturbative
contributions. It makes little sense to focus first on stabilizing moduli
from the flux superpotential only  which is actually ``too stable'' to lead
to relevant phenomenology once the condensate contributions are added.

\section{Gaugino condensation, stabilization of moduli and supersymmetry
breaking}\label{ftsgs}

Nonperturbative phenomena do not necessarily break supersymmetry,
independently of their effect on the stabilization of the moduli
and on the positivity of the potential. To illustrate these
statements, we will proceed by analyzing some examples, insisting
on the role of the various terms that can appear in the
superpotential.

\subsection{Some properties of the scalar potential stationary points}
\label{secVmin}

The analysis of the non-positive scalar potential depending on
seven complex fields is a difficult task when supersymmetry
breaks. We will use in this section some properties of the
potential which follow from our particular forms of the K\"ahler
potential and of the superpotential.

In a general supergravity theory with K\"ahler potential $K = - \sum_j\ln(
Z_j + \ov Z_j)$, supersymmetry is spontaneously broken if the equations
 \beq \label{Fjis} F_j  \equiv W - (Z_j + \ov Z_j )
 W_j =0
 \eeq
cannot be solved for all scalar fields $Z_j$ (and with $\Re Z_j>0$). In this case, our
first goal is to find stationary points where supersymmetry breaks
in Minkowski space:
 \beq \langle V \rangle =
 0 \, , \qquad\qquad \langle W \rangle \ne 0 \,.
 \eeq
A stationary point of the scalar potential is a solution of the
equation $\partial_j V =0$, $\forall j$, which explicitly reads:
 \beq \label{Vmin} 0 = {\rm e}^{-K}V K_j  - \ov W_j F_j - 3
 W_j\ov W + \sum_{i{\rm \; with \; }i\ne j} \left[ W_j - (Z_i+\ov Z_i)W_{ij} \right]
 \ov F_i -( Z_j + \ov Z_j ) W_{jj} \ov F_j
 \eeq
for each scalar field $Z_j$. The first term vanishes at a
Minkowski point and the second derivative $W_{jj}$ only exists for
the modulus appearing in the exponential gaugino condensate term.

Suppose now that $\langle W_j \rangle = 0$ for a certain modulus.
Then $\langle F_j \rangle = \langle W \rangle \ne 0$ and the
contribution of the field to $\langle V \rangle$ cancels one unit
of the negative term $-3\langle W\ov W\rangle$. We then consider the case
where the scalar fields split in two categories,
with either $\langle W_j\rangle =0$ and $\langle F_j\rangle = \langle W \rangle \ne
0$, or with $\langle F_j\rangle =0$.  Supersymmetry breaking is controlled
by the first category only. The Minkowski condition, $\langle V\rangle =0$
implies then that this category contains precisely three fields.

For a field such that $\langle W_j \rangle = 0$, the stationarity condition (\ref{Vmin})
simplifies to\footnote{This assumption will not be used for the
modulus controlling the gaugino condensate.}
 \beq 0 = \sum_{i{\rm \; with \; } i\ne
 j} (Z_i + \ov Z_i)\,  W_{ij}\,  \ov F_i  .\label{Vminbr}
 \eeq
Only directions with broken supersymmetry contribute to the
summation, since for the others $\langle F_i\rangle =0$. For those
which contribute, $\langle F_i\rangle =\langle W \rangle$ and
therefore Eq.~(\ref{Vminbr}) reduces to
 \beq 0 = \sum_{i{\rm \; with \; } i\ne
 j} W_{ij}\,  \Re Z_i  \, ,
 \label{VminbrFneq0}
 \eeq
where the summation includes only fields $Z_i$ such
that $\langle W_i \rangle = 0$. Notice that this equation is
trivially satisfied whenever fields such that $\langle W_j\rangle
=0$ couple in the superpotential only to directions where
supersymmetry does not break. In this case, $ W_{ij}$ in
Eq.~(\ref{VminbrFneq0}) vanishes since $W_{ij} \ne 0
\Longleftrightarrow \langle F_i \rangle =0$.

Consider now a modulus $Z_j$ for which $\langle F_j \rangle
= 0$. Equation (\ref{Vmin}) for this
field becomes
 \beq 0 = - 3 W_j\ov W + \sum_{i{\rm \; with \; } i\ne j} \biggl[
 W_j - (Z_i+\ov Z_i)W_{ij} \biggr] \ov F_i.\label{Vmin2}
 \eeq
The sum runs over fields for which $\langle F_i\rangle \ne 0$. For
those, $\langle W_i \rangle = 0$ and therefore $\langle F_i
\rangle = \langle W \rangle $. Since there are three such fields,
Eq.~(\ref{Vmin2}) is equivalent to
 \beq 0 = \sum_{i{\rm \; with \; } i\ne j} W_{ij}\,  \Re Z_i \, ,
 \label{VminF=0}
 \eeq
with a summation over all fields $Z_i$ such that $\langle W_i
\rangle = 0$. Hence, both conditions can be summarized by the
seven conditions \beq\label{Vminall} 0 = \sum_{i{\rm \; with \; }
i\ne j} \langle W_{ij}\Re Z_i \rangle , \eeq with a summation
restricted over moduli which break supersymmetry.

In the examples studied below, the structure of the superpotential
gives indeed a partition between moduli for which $\langle
W_j\rangle=0$ and supersymmetry breaks, $\langle F_j\rangle \ne0$,
and moduli for which supersymmetry does not break $\langle
F_j\rangle =0$. For the first class, the stationarity of the
potential reduces to the simple equation $W_j=0$ while for the
second class, it reduces to Eq.~(\ref{VminF=0}).

\subsection{Supersymmetry breaking independent of the gaugino
condensation}\label{sbsg}

Let us start with the case where the gaugino condensate does not
break supersymmetry. The introduction of fluxes can indeed
stabilize some of the moduli in flat space without inducing
supersymmetry breaking. To be concrete, let us consider a superpotential
with ``supersymmetric mass terms" only:
\begin{eqnarray}
     W_{\rm susy} &=& A (U_1-U_2) (T_1 - T_2) + B (U_1 + U_2 - 2 U_3) ( T_1 + T_2 - 2
     T_3)\nonumber \\  &+&  C (U_1-U_2)( T_1 + T_2 - 2 T_3) + D (T_1 - T_2)
     (U_1 + U_2 - 2
     U_3).
\end{eqnarray}
This superpotential is created by geometrical fluxes, either in
heterotic or in type IIA. It can be directly generated at the
string level using freely acting orbifold constructions.
It selects four (complex) directions in the seven-dimensional space of the moduli
fields and minimizing the potential tends to cancel the fields in these four
directions.
Supersymmetry is not broken and cancellation of auxiliary
fields\footnote{The conditions $W_i\left(Z_i+\ov Z_i\right) = W$
for every scalar field $Z_i$.} fixes $U_A =U$ and $T_A =T$. There
are then flat directions and $W_{\rm susy} = 0$ at the minimum.
Further supersymmetric mass terms are those which mix the $T_A$
and $U_A$ moduli in the minimization conditions:
\begin{equation}
     (U_1-T_1)(U_2-U_3)\quad{\rm or} \quad (U_1-T_1)(T_2-T_3).
\end{equation}
The first term is present only in heterotic with geometrical and
$H_3$ fluxes, while the second term appears only in type IIA,
when geometrical and $F_2$ fluxes are switched on. We could also
consider
\begin{equation}
     (U_1-T_1)(U_2-T_2)
\end{equation}
which exists as a four-dimensional $N=4$ gauging but cannot be
obtained from a ten-dimensional field theory by the
Scherk--Schwarz mechanism, although it can be realized in freely
acting asymmetric heterotic orbifold constructions \cite{ferm}.

In any case, since the superpotential is homogeneous, the
normalizations of the $T_A$ and $U_A$ moduli are not fixed and at
least one modulus direction remains flat. These supersymmetric
mass terms are generic in the sense that they may fix some of the
moduli without modifying the minimum of the potential and without
breaking supersymmetry. We will use those terms in the following
as building blocks, and add nonperturbative contributions.

Our first example is the following:
 \beq
 \label{susy1}
\begin{array}{rcl}
 W_{\rm susy} &=& A (U_1-U_2) (T_1 - T_2) +
 B (U_1 + U_2 - 2 U_3 ) ( T_1 + T_2 - 2 T_3)
\crbig && + (T_1+T_2-2T_3) \, w(S) ,
\end{array}
\eeq
with $w(S)$ as given in Eq.~(\ref{ftW}). The condensate term
$(T_1 + T_2 - 2 T_3) w(S)$ can be understood from the general form
of the $N=4$ superpotential, Eq.~(\ref{superpot}). This cubic
expression also produces in the superpotential terms proportional
to
$$
f_{IJK}\epsilon_{ABC} \Phi_A^I Z_B^J Z_C^K
$$
where $\Phi_A^I$ is linear in the main moduli $T_A$ and $U_A$ while
the $Z_B^J$ and $Z_C^K$
are charged under the confining gauge group. As discussed earlier, condensation
induces for these charged contributions an exponential dependence on the gauge
coupling field of the confining gauge group, say $S$,
as in the last term of superpotential (\ref{susy1}).

The presence of $w(S)$ leaves $U_1 = U_2 = U$ but we now have
$U_3=U  + w(S)/2B $. The above conclusions about supersymmetry
remain however unchanged: supersymmetry is unbroken and the
gravitino is massless in flat background.

The previous example can be modified by the addition of further
flux terms which break supersymmetry. Consider instead, in the
heterotic or type IIA \beq \label{het1} W_{\rm total} = W_{\rm
susy} + W_{\rm break} \eeq with $W_{\rm susy}$ as in
Eq.~(\ref{susy1}) and \be
 W_{\rm break}= R\,(T_1 U_1 + T_2 U_2).
 \ee
The term $W_{\rm break}$ breaks supersymmetry \emph{even in the
absence} of $w(S)$. The scalar potential has a minimum with
real $T_A$, $U_A$ and $T_A = T$, $U_1 = U_2 = U$, $U_3 = U +
w(S)/2B$. The potential vanishes along the flat directions $S$,
$T$ and $U$. The goldstino field is a combination of the fermionic
partners of $S, T_3$ and $U_3$. There is an ``effective'' no-scale
structure: since $W_i=0$ in these three directions, their
corresponding contributions to the potential cancel the
gravitational contribution $-3m_{3/2}^2$. Thus supersymmetry is
broken in flat space--time with
$$
m_{3/2}^2 = {|R|^2\over 32 \, ST_3U_3}
$$
and the presence of the nonperturbative term $w(S)$ only acts as a
small perturbation on supersymmetry breaking induced by the
modulus-dependent contribution $W_{\rm break}$. It however
explicitly appears in mass terms.

Keeping the same mass terms given in Eq.~(\ref{susy1}) but
choosing a different breaking term
\beq
 W_{\rm
 break}= R\,(T_1 S + L U_3 T_3),
 \eeq
induced by the geometrical
fluxes in type IIA, fixes the moduli after minimization to
$T_A=T$, $S=L U_3$, $U_3= U + w(S)/2B$. Now the goldstino direction
is a combination of $U_1, U_2$ and $T_2$ and in this example, the
$S$-direction does not break supersymmetry.

As another example in the framework of type IIA, keep the same
mass terms but replace $W_{\rm break}$ in Eq.~(\ref{het1}) by
\begin{eqnarray}
W_{\rm break}= R\left[S (T_1 + T_2 + T_3) + T_1 T_2 + T_2
T_3+T_3T_1\right] + i M (L S + T_1 T_2 T_3 ).
\end{eqnarray}
The terms proportional to $ST_A$ come from the geometrical fluxes,
those proportional to $T_AT_B$ are generated by  the R-R two-form
fluxes $F_2$, the $T_1 T_2 T_3$ term by the zero-form flux $F_0$
and the term linear to $S$ originates from the NS three-form
fluxes $H_3$. The minimization conditions now give: $S=T_A =T=
\sqrt{L}$, $U_3 = U + w(S)/2$. The supersymmetry is broken and the
potential is again (locally) semi-positive-definite, compatible with flat
space--time. The goldstino superfield is a combination of $U_1,
U_2$ and $U_3$ superfields. In this example as well, the
$S$-direction does not break supersymmetry.

Similar situations are met in type IIB. Depending on the  kind of
confining gauge group ($D_3$-branes versus  $D_7$-branes) the
gauge coupling constant can be either $S$ or one of the $T_A$
moduli, say $T_1$. Let us consider for instance the following
superpotential, valid in the presence of $D_3$-branes:
 \be\label{IIB2}
 W_{\rm total}= A(U_1-U_2)(S - U_3 - w(S)) + W_{\rm break}
 \ee
with
 \be W_{\rm break} = iB(L S + U_1U_2U_3).\label{IIB2br}
 \ee
The terms proportional to $S$ originate from $H_3$; the others
from $F_3$.

In the absence of $W_{\rm break}$, supersymmetry remains unbroken
even when NS-NS and R-R three-form fluxes are on, as in example
(\ref{IIB2}). With the actual $W_{\rm break}$
[Eq.~(\ref{IIB2br})], supersymmetry is broken with
semi-positive-definite potential as in all previous examples. The
minimization of the potential leads to:
 \be U_1=U_2 =U, \qquad  S =
 U_3+ w(S),\qquad LS=U^2 [ S-w(S) ],
 \ee
where we assumed $U$ and $S$ real. The goldstino superfield is
defined by the $T_A$'s,  which cancel the negative contribution of
the potential and give rise to a no-scale structure, even in the
presence of nonperturbative terms. Finally
\begin{equation}
  W_{\rm min}=2iBU^2 [ S-w(S) ] =2iBLS.
\end{equation}

We can examine the other case, namely when the gauge coupling
constant is $T_1$. We must then replace $w(S)$ by $w(T_1)$,
 \be
 W_{\rm total}= A(U_1-U_2)(S - U_3 - w(T_1)) + W_{\rm
 break},
 \ee
and choose the same (or a different) breaking term. Choosing
$W_{\rm break}$ as in Eq.~(\ref{IIB2br}), the minimization of the
potential leads to:
  \be U_1=U_2 =U, \qquad  S = U_3+ w(T_1),\qquad
 LS=U^2 [ S-w(T_1) ]. \ee
The goldstino superfield is defined again by the $T_A$'s and the
potential has a no-scale structure as before, even in the presence
of the $T_1$--dependent nonperturbative terms.

Notice incidentally that $\mu^3$ in $w(S {\ \rm or\ } T_A)$
[Eq. ~(\ref{ftW})] can generally depend on $Z_i$ moduli. This dependence
has no consequence on the various properties discussed previously.

In the examples we have exhibited so far, the role of the gaugino
condensate was important but not crucial for the supersymmetry
breaking. An explicit breaking term, generated by a specific
combination of fluxes, had to be superimposed to the mass terms,
in order for the supersymmetry to be broken, independently of the
presence of the condensate.

This situation is not generic, and we will now describe another
class of examples, where supersymmetry breaking is triggered
by the gaugino condensate.

\subsection{Gaugino-induced supersymmetry breaking in type IIA}\label{sbag}

One generic feature of the breaking mechanism described in Sec.
\ref{sbsg} is that the nonperturbative contribution to the
gravitino mass  is negligible. On the contrary, we will now
analyze situations where the gaugino condensate breaks
supersymmetry, and $m_{3/2}$  turns out to be related to the
gaugino scale $w(S)$ only.

The generic form of the superpotential is again
 \beq
 W= W_{\rm susy} + \mu^3(Z_i) \, {\rm e}^{-S}.
 \eeq
The nonperturbative term can be used to induce supersymmetry
breaking and thus it fully contributes to the effective theory at
the vacuum. In other words, in contrast to previous examples,
$\mu^3(Z_i)$ will not vanish at the minimum.

In order to fix the ideas we will examine this situation in detail
in examples of type IIA orientifolds. The structure of those
examples shows how generic is the procedure in the present
framework.

Consider the superpotential
\begin{equation}
\label{WIIAsb} W= \left(T_1 - T_2\right)\left(- U_1+ U_2  - T_3  +
2S \right) + \left(U_1 T_3 -L\right)w(S),
\end{equation}
which is generated by geometric and $F_2$ fluxes. It falls in the class
mentioned in Sec. \ref{secVmin}, where there is a partition
between directions which break supersymmetry (here $T_1$ and
$T_2$) and directions which preserve supersymmetry ($T_3, U_1,
U_2$ and $S$).

The  requirement $\langle W_{T_1} \rangle = \langle W _{T_2}
\rangle= 0$ ensures $\langle V \rangle =0$ since $W$ is
independent of $U_3$, and the resulting supersymmetry-breaking
condition reads
\begin{equation}
\label{sb1}   - U_1 + U_2- T_3 +2 S =0.
\end{equation}
The vanishing of the $F$-auxiliary fields in the directions
$T_3,U_1,U_2$ and $S$ leads to the following equations:
\begin{eqnarray}
\xi\left(\ov U_1 + U_2 - T_3 + 2  S\right) - \left( \ov
 U_1 T_3 + L\right)w(S)&=&0, \label{co1IIAsb1}\\
\xi\left(- U_1  -\ov U_2 - T_3 + 2  S\right) + \left(
 U_1 T_3 - L\right)w(S)
&=&0,\label{co2IIAsb1} \\
\xi\left(- U_1  + U_2 + \ov T_3 + 2  S\right) - \left(
 U_1 \ov T_3 + L\right)w(S)
&=&0, \label{co3IIAsb1}\\
\xi\left( - U_1  + U_2 - T_3 - 2  \ov S\right) + \left(
 U_1 T_3 - L\right)\left(
 1 + S + \ov S\right)w(S)
&=&0,\label{co4IIAsb1}
\end{eqnarray}
where we have introduced
 \begin{equation}\label{xi}
 \xi \equiv T_1 - T_2.
 \end{equation}
The minimization condition (\ref{VminF=0}) reads here
\begin{equation}\label{minxi}
\Re  \xi =0.
\end{equation}
Equations (\ref{sb1})--(\ref{minxi}) must be solved for $\xi, T_3,
U_1, U_2$ and $S$. Combining Eqs.~(\ref{co1IIAsb1}) and
(\ref{co3IIAsb1}), one concludes that $T_3 = U_1$. A similar
combination of Eqs.~(\ref{co1IIAsb1}) and (\ref{co2IIAsb1}) shows
that these moduli must be chosen real: $T_3 = U_1 = t$. The
requirement (\ref{minxi}) can be fulfilled by adjusting
appropriately the imaginary part of the $S$ field: $S = s - i
\frac{\pi}{2}+3i\varphi_\mu$ ($\mu = |\mu|\exp i\varphi_\mu$).
This implies through Eq.~(\ref{sb1}) that $U_2 = u + i
\pi-6i\varphi_\mu$.

Finally, the equations for $t,u$ and $s$ are (\ref{sb1}), a
combination of (\ref{co1IIAsb1}) and (\ref{co2IIAsb1}) as well as
a combination of (\ref{sb1}), (\ref{co1IIAsb1}) and
(\ref{co4IIAsb1}), which read:
 \begin{eqnarray}
 u+2(s-t)
 &=&0, \label{1IIAsb1}\\
  {t}\left(t^2-L\right)- {u}\left(t^2+L\right)
 &=&0, \label{2IIAsb1}\\
 t^5+ 2 L t^3- 4 L t^2- 3 L^2 t-4 L^2 &=&0. \label{3IIAsb1}
 \end{eqnarray}
Furthermore $\xi$ will be given by any of the equations
(\ref{co1IIAsb1}) to (\ref{co4IIAsb1}), once the other moduli are
fixed:
\begin{equation} \label{T12sb1}
\xi =  \frac{t^2 + L}{u + 2 s}w.
\end{equation}
Similarly, the gravitino mass will read:
 \beq {\rm
e}^{-K/2} \, m_{3/2} =  \langle W \rangle =\left( t^2 - L\right)
w(s). \eeq

From Eq.~(\ref{3IIAsb1}), assuming $L>0$ we obtain:
\begin{equation} \label{tLsb1}
L = t^2 \frac{t-2+2\sqrt{t^2 + 1}}{3t+4};
\end{equation}
the choice for the sign of the square root is the only one
compatible with positivity of all moduli. Assuming $L$ large, the
leading and sub-leading behaviour for $t$ is
\begin{equation} \label{leadtLsb1}
t= \sqrt{L} + 1 + \mathcal{O}\left(\frac{1}{\sqrt{L}}\right).
\end{equation}
Equation (\ref{2IIAsb1}) gives thus
\begin{equation} \label{leaduLsb1}
u= 1 + \mathcal{O}\left(\frac{1}{L}\right),
\end{equation}
whereas from (\ref{1IIAsb1}) we get:
\begin{equation} \label{leadsLsb1}
s= \sqrt{L} + \frac{1}{2} +
\mathcal{O}\left(\frac{1}{\sqrt{L}}\right).
\end{equation}
As advertised previously, supersymmetry breaking is measured by
 \beq {\rm Im} \, \xi \approx \sqrt{L}\, |\mu|^3 \, {\rm
 e}^{-\sqrt{L}} \eeq
and the gravitino mass scales as
 \beq {\rm
 e}^{-K/2} \, m_{3/2} \approx 2i \sqrt{L}\, |\mu|^3 \, {\rm
 e}^{-\sqrt{L}}. \eeq

In the example at hand, the gaugino condensate is entirely
responsible for the breaking of supersymmetry, as shown by the
last two equations. Furthermore, the fluxes generating the
superpotential (\ref{WIIAsb}) are not fine-tuned, and solutions
for the moduli exist generically.

In fact, many examples of the above type can be found in type IIA,
which are generated by fluxes originating from string theory
compactifications. Consistent superpotentials with similar
gaugino-condensation-induced supersymmetry breaking also exist,
which are valid as supergravities without stringy origin. This
happens for example with
\begin{equation}
\label{WIIAsb2} W= \left(T_1 - T_2\right)\left(U_1 + U_2 + b U_3
+g U_1 U_2 \left(U_3 -i \pi\right)- 2b S \right) + \left(  U_1 U_2
- L\right)w(S),
\end{equation}
where the cubic term cannot be obtained by switching on fluxes in
type IIA string compactifications. Although this superpotential
turns out to possess the various features advertised so far, we
will not further elaborate on it.

\subsection{Gaugino-induced supersymmetry breaking in heterotic}\label{shet}

Although type II and heterotic compactifications are quite similar
when supersymmetry breaking is mostly due to fluxes [see Sec.
(\ref{sbsg})], they turn out to be drastically different in cases
where the breaking of supersymmetry is induced by a gaugino
condensate. The reason boils down to the absence of $S$
contributions to the superpotential, directly originating from
fluxes.

Consider for concreteness a superpotential of the type
\begin{equation}\label{Whet}
 W= {\hat A} U_1 + {\hat B} U_2 +{\hat C} U_3 + {\hat D} U_4,
\end{equation}
where $U_4 = U_1 U_1 U_3$. This superpotential is odd in the
$U_i$'s and captures most of the heterotic compactifications
considered in the present work, with a gaugino condensate. We have
introduced the following functions of $T_1,T_2$ and $S$:
\begin{eqnarray}
 {\hat A}&=& \Big[\alpha + \alpha' w(S)\Big]\xi + Aw(S), \label{WhetA}\\
 {\hat B}&=& \Big[\beta + \beta' w(S)\Big]\xi + Bw(S), \label{WhetB}\\
 {\hat C}&=& \Big[\gamma + \gamma' w(S)\Big]\xi + Cw(S), \label{WhetC}\\
 {\hat D}&=& \Big[\delta + \delta' w(S)\Big]\xi + Dw(S),\label{WhetD}
\end{eqnarray}
where $\xi = T_1-T_2$ as defined in (\ref{xi}) and $w(S)$ in
(\ref{ftW}).

The minimization condition (\ref{VminF=0}) reads $\Re \xi =0$, as
in the examples of type IIA (Sec. \ref{sbag}). We will therefore
choose $S= s -i{\pi/2}+3i\varphi_\mu$ and $U_i= u_i$ real.
Everything is consistent provided $\alpha,\beta,\gamma,\delta$ and
$A,B,C,D$ are real and $\alpha',\beta',\gamma',\delta'$ are
imaginary.

The no-scale requirement $\langle V \rangle =0$ is fulfilled
provided $\langle W_{T_1} \rangle = \langle W _{T_2} \rangle= 0$
($W$ is independent of $T_3$). The corresponding condition reads:
\begin{equation}\label{sbhet}
  (\alpha + \alpha' w)u_1 +(\beta + \beta' w)u_2 + (\gamma +
\gamma'w)u_3 +(\delta + \delta' w)u_4=0,
\end{equation}
whereas the vanishing of the $U_A$--auxiliary fields leads to
\begin{eqnarray}
 -{\hat A} u_1 + {\hat B} u_2 +{\hat C} u_3 -{\hat D} u_4&=&0 , \label{hetu1}\\
  {\hat A} u_1 - {\hat B} u_2 +{\hat C} u_3 - {\hat D} u_4&=&0 , \label{hetu2}\\
  {\hat A} u_1 + {\hat B} u_2 -{\hat C} u_3 - {\hat D} u_4&=&0 . \label{hetu3}
\end{eqnarray}
These equations can be solved. They indeed imply that
\begin{equation}\label{hetu}
  {\hat A} u_1 ={\hat B} u_2 ={\hat C} u_3 =  {\hat D}
  u_1u_2u_3,
\end{equation}
which allows therefore to express $u_1, u_2, u_3$ in terms of
$\xi, s$:
\begin{equation}\label{hetusol}
 u_1=\sqrt{\hat B\hat C\over \hat A \hat D}\ ,\ \
 u_2=\sqrt{\hat A \hat C\over\hat B \hat D}\ ,\ \
 u_3=\sqrt{\hat A \hat B\over\hat C \hat D}\ ,\ \
 u_4=\sqrt{\hat A \hat B\hat C\over \hat D^3}.
\end{equation}
The fields $\xi$ and $s$ are in turn determined by Eq.~(\ref{sbhet})
and the equation for the $S$-auxiliary field:
\begin{equation}\label{hets}
  W+ 2s \Big[\left(\alpha' \xi + A
\right)u_1 + \left(\beta' \xi + B\right)u_2
+\left(\gamma'\xi+C\right)u_3 + \left(\delta' \xi+D
\right)u_4\Big]w=0,
\end{equation}
which can be further simplified by using Eqs.~(\ref{sbhet}) and
(\ref{hetusol}):
\begin{equation}\label{hetssol}
  {2 \over s}=-4-\left( {\alpha '\over \hat A} + {\beta '\over \hat
B}+{\gamma '\over \hat C}+ {\delta '\over \hat D}\right)\xi w.
\end{equation}

In order to solve for $\xi$ and $s$, we will introduce a set of
intermediate imaginary quantities $\lambda_i$, defined by
\begin{equation}\label{lamdef}
  {\hat A}=\lambda_1\xi w\ , \ \ {\hat B}= \lambda_2\xi w\ ,\ \ {\hat
C}=\lambda_3\xi w\ , \ \ {\hat D}=\lambda_4 \xi w.
\end{equation}
Using Eqs.~(\ref{WhetA})--(\ref{WhetD}), these parameters are
expressed in terms of $\xi$ and $s$:
\begin{eqnarray}
 \alpha \xi + Aw +(\alpha'-\lambda_1)\xi w&=& 0, \label{lamxiA}\\
 \beta \xi +Bw+(\beta'-\lambda_2)\xi w&=& 0, \label{lamxiB}\\
 \gamma\xi +Cw +(\gamma'-\lambda_3)\xi w&=& 0, \label{lamxiC}\\
 \delta\xi +Dw + (\delta'-\lambda_4)\xi w&=& 0.\label{lamxiD}
\end{eqnarray}
The $\lambda_i$'s allow to express all $u_i$'s as
\begin{equation}\label{lamu}
u_1=\sqrt{\lambda_2\lambda_3\over \lambda_1\lambda_4}\ ,\ \
u_2=\sqrt{\lambda_1\lambda_3\over \lambda_2\lambda_4}\ ,\ \
u_3=\sqrt{\lambda_1\lambda_2\over \lambda_3\lambda_4}\ ,\ \
u_4=\sqrt{\lambda_1\lambda_2\lambda_3\over \lambda_4^3}.
\end{equation}
These expressions show in particular that an even number of
$i\lambda_i$'s can be negative. Finally, Eqs.~(\ref{sbhet}) and
(\ref{hetssol}) read:
\begin{equation}\label{lamsbhet}
  {\alpha + \alpha' w\over \lambda_1}+ {\beta+\beta'w \over
\lambda_2}+{\gamma+\gamma'w \over \lambda_3}+{\delta+\delta'w
\over \lambda_4}=0
\end{equation}
and
\begin{equation}\label{lamhetssol}
  {2 \over s}=-4-\left( {\alpha '\over \lambda_1} + {\beta '\over
\lambda_2}+{\gamma '\over \lambda_3}+ {\delta '\over
\lambda_4}\right).
\end{equation}
These are the final equations for determining $\xi$ and $s$. We
will now proceed and show that solutions with large and positive
$s$, together with exponentially small $\xi$ ({\it i.e.} perturbative
regime and small supersymmetry breaking) do indeed exist under
minor and natural assumptions on the fluxes (coefficients
$\alpha,\beta,\gamma,\delta,\alpha',\beta',\gamma',\delta'$ and
$A,B,C,D$).

For simplicity, we restrict ourselves to the plane-symmetric
situation, where
\begin{equation}
\alpha = \beta = \gamma\ , \ \  \alpha' = \beta' = \gamma'\ , \ \
A = B = C,
\end{equation}
which imply that
\begin{equation}\label{lamups}
\lambda_1=\lambda_2=\lambda_3\equiv\lambda \ \ {\rm and} \ \  u_1
= u_2 = u_3 = \sqrt{\lambda\over \lambda_4} \equiv u .
\end{equation}
The set of equations we must solve is therefore (\ref{lamsbhet})
and (\ref{lamhetssol}) together with those defining $\lambda$ and
$\lambda_4$, namely (\ref{lamxiA}) and (\ref{lamxiD}). We can
eliminate $\xi$ though the latter equations and use
(\ref{lamsbhet}) to express $\lambda$ and $\lambda_4$ as functions
of $s$ only:
\begin{eqnarray}
{1\over \lambda}&=&{w\over 3(\alpha+\alpha'w)}{3D\alpha+A\delta +
(3\alpha' D +A\delta')w
\over D\alpha-A\delta +(D\alpha'-A\delta') w }, \label{lam}\crbig
{1\over \lambda_4}&=&-{w\over (\delta+\delta'w)}{3D\alpha+A\delta
+ (3\alpha' D +A\delta')w \over D\alpha-A\delta
+(D\alpha'-A\delta') w }. \label{lam4}
\end{eqnarray}
We can now express $\xi=\xi(s)$ using either (\ref{lamxiA}) or
(\ref{lamxiD}) together with (\ref{lam}) and (\ref{lam4}):
\begin{equation}\label{xieq}
4\xi=-{3D\alpha+A\delta+(3D\alpha'+A\delta')w \over
(\alpha+\alpha'w)(\delta+\delta'w)}\, w,
\end{equation}
whereas the central equation for the determination of $s$ will be
given by (\ref{lamhetssol}):
\begin{equation}\label{seq}
{2\over s}=-4-{(\alpha'\delta-\delta'\alpha)w\over
(\alpha+\alpha'w)(\delta+\delta'w)}{3D\alpha+A\delta + (3\alpha' D
+A\delta')w \over
 D\alpha-A\delta+(D\alpha'-A\delta')w
}.
\end{equation}

We would like now to show that Eq.~(\ref{seq}) admits physically
acceptable solutions for $s$, provided that the fluxes
$\alpha,\delta, \alpha', \delta',A,D$ are large while their ratios
are of order unity. If this requirement is fulfilled we can define
a variable $\rho$ (real function of $s$) as
\begin{equation}\label{rho}
 \rho =i{D\alpha-A\delta\over D\alpha w  },
\end{equation}
which can be consistently taken to be of order one since $w$ is
small and $D\alpha/A\delta$ of order one.

Under these assumptions, we can perform an expansion in powers of
$w$ for all quantities. We find the following dominant
contributions:
\begin{equation}\label{xiapprox}
\xi\approx-{D\over \delta}~w
\end{equation}
and
\begin{equation}\label{sapprox}
 {1\over 2s} \approx -1-{D\over\delta}
 {\alpha'\delta-\delta'\alpha\over D\alpha'-A\delta' -iD\alpha \rho }.
\end{equation}
For further simplification we specialize to
\begin{equation}\label{spe}
\alpha'=i\alpha\ , \ \ \delta'=-i\delta.
\end{equation}
Equations (\ref{lam}) and (\ref{lam4}) give, at dominant order,
\begin{equation}\label{sapproxspe}
{1\over 2s}\approx{4-\rho \over \rho-2} \quad\Longleftrightarrow\quad
\rho\approx 2 {4s+1\over 2s+1}.
\end{equation}
Notice also that from Eq.~(\ref{lamups}) we obtain
\begin{equation}\label{uapproxspe}
u\approx  \sqrt{-3\alpha \over\delta}.
\end{equation}
Inserting the definition of $\rho$ [Eq.~(\ref{rho})] into
(\ref{sapproxspe}) we obtain the following equation for $s$, at
dominant order:
\begin{equation}\label{sapproxspes}
  {A\delta-D\alpha\over D\alpha }\approx2i{4s+1\over 2s+1}w(s).
\end{equation}
As advertised previously, this equation is compatible with large
values of $s$. In that regime it further simplifies:
\begin{equation}\label{sapproxspess}
s\approx\log\left({{4|\mu|^3 D\alpha\over D\alpha-A\delta
}}\right)-{1\over 4\log\left({{4|\mu|^3 D\alpha\over
D\alpha-A\delta }}\right)}
\end{equation}

We can finally determine the gravitino mass. Using
Eqs.~(\ref{Whet}),  (\ref{lamdef}) and (\ref{lamu}), we obtain:
\begin{eqnarray}
{\rm e }^{-K/2}m_{3/2}&=& \langle W\rangle
\nonumber \\
&=& 4 u^3 \lambda_4  \xi w = 4 \sqrt{\lambda^3\over \lambda_4} \xi
w
\nonumber \\
&=&-\frac{A\delta' - D\alpha' + \frac{A \delta - D\alpha
}{w}}{\alpha+\alpha'w}\left(-3\frac{\alpha+\alpha'w}{\delta+\delta'w}\right)^{3/2}w^2
\end{eqnarray}
for generic plane-symmetric situations. In the special case
captured by (\ref{spe}) and within the above approximations, the
result is
\begin{equation}
{\rm e }^{-K/2}m_{3/2} \approx i4D\left(-
\frac{3\alpha}{\delta}\right) ^{3/2}\frac{s}{2s+1} w^2,
\end{equation}
with $s$ given in Eq.~(\ref{sapproxspess}). The gravitino mass scales
as $w^2$ instead of $w$ like in type IIA (Sec. \ref{sbag}). This
is due to the absence of flux-induced $S$-term in the heterotic
superpotential.

\section{Conclusions}

Understanding the effect of nonperturbative corrections in the
presence of fluxes is an important issue. As already stressed,
this can shed light on the nature of the vacuum and the
stabilization of moduli, and turns out to be a valuable tool for
circumventing the runaway behaviour of moduli or the fine-tuning
problem.

In the present paper, we have addressed these questions through a
selection of examples, using the effective supergravity analysis
as obtained from type II orientifolds and heterotic string
compactifications. The important outcome of this analysis is that
the pathological behaviour of the vacuum in the presence of fluxes
with nonperturbative corrections \emph{is not} a generic property
of the $N=1$ effective supergravity, with or without spontaneously
broken supersymmetry: the usual caveats quoted previously can be
avoided with a suitable combination of flux and nonperturbative
contributions.

We would like to emphasize the importance of analyzing  the entire
superpotential, including both flux-induced perturbative and
nonperturbative contributions. Although nonperturbative
corrections do not necessarily trigger supersymmetry breaking,
they can alter various terms and must therefore be taken into
account at a very early stage. Most importantly, they can severely
change the picture of moduli stabilization drawn by fluxes only.
This has been illustrated in several examples in Sec. \ref{ftsgs}.

The examples that we have investigated enable us to conclude that
at least \emph{three} scaling behaviours of the gravitino mass
exist:
\begin{enumerate}
\item In models where the nonperturbative corrections do not
  trigger the supersymmetry breaking -- they modify the mass terms though
  -- the gravitino mass scales like:
  $$m_{3/2} = c\,/ \sqrt{V} , $$
  where $V = \exp K$ is the volume of the moduli space and $c$
  is related to the flux numbers. In this case, any mass hierarchy
  strongly relies on the volume $V$.

\item  In type II models where the nonperturbative contributions to the
  superpotential induce the supersymmetry breaking, the gravitino mass
  is controlled by the nonperturbative superpotential $w(S)$. We
  have given examples where
  \begin{equation}\label{IIAhier}
  m_{3/2} = c\,w(S) \,/ \sqrt{V},
  \end{equation}
  as commonly expected from nonperturbative supersymmetry breaking.
  In this case, a mass hierarchy can be created irrespectively of
  the size of the volume of the moduli space.
\item

A \emph{third} scaling behaviour actually appears in the framework
of heterotic string when supersymmetry breaking is directly
induced by the gaugino condensate. There, the gravitino mass is
more unusual but still generic:
  \beq
  m_{3/2} = c\,w(S)^2\, / \sqrt{V}.
  \eeq
This creates a mass hierarchy stronger than in type IIA
[Eq.~(\ref{IIAhier})]. The heterotic realizations under consideration
are actually quite generic despite the fact that the ratios
between flux coefficients are required to be of order one, while
the coefficients themselves are large.
\end{enumerate}
The last case shows that the analysis of heterotic models with
supersymmetry breaking induced by non-perturbative effects
deserves a more systematic investigation. To be complete, the
latter might necessitate the inclusion of glueball terms of the
type $\langle FF \rangle$ in the superpotential, together with the
gaugino $\langle \lambda\lambda \rangle$ ones. It should shed
light on the validity of various gaugino-condensate-induced
supersymmetry breaking scenarios advertised in the literature but
not captured by our systematic analysis.

One should finally stress that the analysis of the $N=1$,
low-energy soft-supersymmetry-breaking terms strongly depends on
the class of model under consideration since their pattern is
mostly controlled by the radiative corrections induced by the
supersymmetry-breaking sector.

\vspace{1.3cm}
\section*{Acknowledgements}

We thank Fabio Zwirner for discussions. J.-P. D. thanks the Ecole
Normale Sup\'erieure and Ecole Polytechnique. C. K. and P. M. P.
thank the Universit\'e de Neuch\^atel for hospitality. This work
was supported in part by the EU under the contracts
MEXT-CT-2003-509661, MRTN-CT-2004-005104, MRTN-CT-2004-503369, by
the Agence Nationale pour la  Recherche, France, contract
05-BLAN-0079-01, and by the Swiss National Science Foundation.


\newpage
\begin{thebibliography}{99}

\bibitem{gaugino1}
  S.~Ferrara, L.~Girardello and H.~P.~Nilles,
  Phys.\ Lett.\ B {\bf 125} (1983) 457.

\bibitem{gaugino2}
  J.-P.~Derendinger, L.~E.~Ib\'a\~nez and H.~P.~Nilles,
  Phys.\ Lett.\ B {\bf 155} (1985) 65;\\
  M.~Dine, R.~Rohm, N.~Seiberg and E.~Witten,
  Phys.\ Lett.\ B {\bf 156} (1985) 55;\\
  C.~Kounnas and M.~Porrati,
  Phys.\ Lett.\ B {\bf 191} (1987) 91.

\bibitem{het}
  E.~Witten,
  Phys.\ Lett.\ B {\bf 155} (1985) 151;\\
  J.-P.~Derendinger, L.~E.~Ib\'a\~nez and H.~P.~Nilles,
  Nucl.\ Phys.\ B {\bf 267} (1986) 365.

\bibitem{LRS}
  D.~L\"ust, S.~Reffert and S.~Stieberger,
  Nucl.\ Phys.\ B {\bf 706} (2005) 3
  [arXiv:hep-th/0406092].

\bibitem{DKPZ}
  J.-P.~Derendinger, C.~Kounnas, P.~M.~Petropoulos and F.~Zwirner,
  Nucl.\ Phys.\ B {\bf 715} (2005) 211
  [arXiv:hep-th/0411276] and
  Fortsch.\ Phys.\  {\bf 53} (2005) 926
  [arXiv:hep-th/0503229].

\bibitem{DKPZlong}
J.-P.~Derendinger, C.~Kounnas, P.~M.~Petropoulos and F.~Zwirner, to
appear.

\bibitem{hetflux}
A.~Strominger, Nucl.\ Phys.\ B {\bf 274} (1986) 253;\\
R.~Rohm and E.~Witten, Annals Phys.\  {\bf 170} (1986) 454; \\
N.~Kaloper and R.~C.~Myers, JHEP {\bf 9905} (1999) 010
[arXiv:hep-th/9901045].

\bibitem{IIAflux}
J.~Polchinski and A.~Strominger, Phys.\ Lett.\ B {\bf 388} (1996)
736 [arXiv:hep-th/9510227];\\
I.~Antoniadis, E.~Gava, K.~S.~Narain and T.~R.~Taylor, Nucl.\ Phys.\
B {\bf 511} (1998) 611 [arXiv:hep-th/9708075];\\
S.~Gukov, C.~Vafa and E.~Witten, Nucl.\ Phys.\ B {\bf 584} (2002)
69 [Erratum-ibid.\ B {\bf 608} (2001) 477] [arXiv:hep-th/9906070];\\
S.~Gukov, Nucl.\ Phys.\ B {\bf 574} (2000) 169
[arXiv:hep-th/9911011].

\bibitem{IIBflux}
J.~Michelson, Nucl.\ Phys.\ B {\bf 495} (1997) 127
[arXiv:hep-th/9610151];\\
K.~Dasgupta, G.~Rajesh and S.~Sethi, JHEP {\bf 9908} (1999) 023
[arXiv:hep-th/9908088];\\
T.~R.~Taylor and C.~Vafa, Phys.\ Lett.\ B {\bf 474} (2000) 130
[arXiv:hep-th/9912152];\\
P.~Mayr, Nucl.\ Phys.\ B {\bf 593} (2001) 99
[arXiv:hep-th/0003198];\\
G.~Curio, A.~Klemm, D.~L\"ust and S.~Theisen, Nucl.\ Phys.\ B {\bf
609} (2001) 3 [arXiv:hep-th/0012213];\\
S.~B.~Giddings, S.~Kachru and J.~Polchinski, Phys.\ Rev.\ D {\bf
66} (2002) 106006 [arXiv:hep-th/0105097];\\
S.~Kachru, M.~B.~Schulz and S.~Trivedi, JHEP {\bf 0310} (2003) 007
[arXiv:hep-th/0201028];\\
A.~R.~Frey and J.~Polchinski, Phys.\ Rev.\ D {\bf 65} (2002) 126009
[arXiv:hep-th/0201029]; \\
S.~Ferrara and M.~Porrati,
Phys.\ Lett.\ B {\bf 545} (2002) 411
[arXiv:hep-th/0207135]; \\
C.~Angelantonj, S.~Ferrara and M.~Trigiante, JHEP {\bf 0310}
(2003) 015 [arXiv:hep-th/0306185] and
Phys.\ Lett.\ B {\bf 582} (2004) 263 [arXiv:hep-th/0310136]; \\
 M.~Grana, T.~W.~Grimm, H.~Jockers and J.~Louis,
  Nucl.\ Phys.\ B {\bf 690}, 21 (2004) [arXiv:hep-th/0312232]; \\
  T.~W.~Grimm and J.~Louis,
  Nucl.\ Phys.\ B {\bf 699}, 387 (2004)
  [arXiv:hep-th/0403067]; \\
 P.~G.~Camara, L.~E.~Ib\'a\~nez and A.~M.~Uranga,
  Nucl.\ Phys.\ B {\bf 708} (2005) 268
  [arXiv:hep-th/0408036];\\
  H.~Jockers and J.~Louis,
  Nucl.\ Phys.\ B {\bf 705}, 167 (2005)
  [arXiv:hep-th/0409098].

\bibitem{CYgen}
  M.~Grana, R.~Minasian, M.~Petrini and A.~Tomasiello,
  JHEP {\bf 0408} (2004) 046
  [arXiv:hep-th/0406137];\\
  C.~M.~Hull and R.~A.~Reid-Edwards,
  arXiv:hep-th/0503114;\\
  T.~House and E.~Palti,
  Phys.\ Rev.\ D {\bf 72} (2005) 026004
  [arXiv:hep-th/0505177];\\
  M.~Grana, J.~Louis and D.~Waldram,
  arXiv:hep-th/0505264.

\bibitem{ferm}
R.~Rohm, Nucl.\ Phys.\ B {\bf 237} (1984) 553;\\
C.~Kounnas and M.~Porrati, Nucl.\ Phys.\ B {\bf 310} (1988) 355;\\
S.~Ferrara, C.~Kounnas, M.~Porrati and F.~Zwirner, Nucl.\ Phys.\ B
{\bf 318} (1989) 75;\\
M.~Porrati and F.~Zwirner, Nucl.\ Phys.\ B {\bf 326} (1989) 162;\\
C.~Kounnas and B.~Rostand, Nucl.\ Phys.\ B {\bf 341} (1990) 641;\\
I.~Antoniadis, Phys.\ Lett.\ B {\bf 246} (1990) 377;\\
I.~Antoniadis and C.~Kounnas, Phys.\ Lett.\ B {\bf 261} (1991)
369;\\
E.~Kiritsis and C.~Kounnas, Nucl.\ Phys.\ B {\bf 503} (1997) 117
[arXiv:hep-th/9703059];\\
I.~Antoniadis, E.~Dudas and A.~Sagnotti, Nucl.\ Phys.\ B {\bf 544}
(1999) 469 [arXiv:hep-th/9807011] and
Phys.\ Lett.\ B {\bf 464} (1999) 38 [arXiv:hep-th/9908023];\\
I.~Antoniadis, J.-P.~Derendinger and C.~Kounnas,
  Nucl.\ Phys.\ B {\bf 551} (1999) 41
  [arXiv:hep-th/9902032], and
  arXiv:hep-th/9908137.

\bibitem{N=4}
A.~Das, Phys.\ Rev.\ D {\bf 15} (1977) 2805;\\
E.~Cremmer and J.~Scherk, Nucl.\ Phys.\ B {\bf 127} (1977) 259;\\
E.~Cremmer, J.~Scherk and S.~Ferrara, Phys.\ Lett.\ B {\bf 74}
(1978) 61;\\
%
A.~H.~Chamseddine, Nucl. \ Phys.\ B {\bf 185} (1981) 403;\\
%
J.-P.~Derendinger and S.~Ferrara, {\it Lectures given at Spring
School of Supergravity and Supersymmetry, Trieste, Italy, Apr
4-14, 1984}, CERN-TH-3903;\\
%
M.~de Roo, Nucl. \ Phys. \ B {\bf 255} (1985) 515 and
Phys.\ Lett.\ B {\bf 156} (1985) 331;\\
E.~Bergshoeff, I.~G.~Koh and E.~Sezgin, Phys.\ Lett.\ B {\bf 155}
(1985) 71;\\
M.~de Roo and P.~Wagemans, Nucl.\ Phys.\ B {\bf 262} (1985) 644
and
Phys.\ Lett.\ B {\bf 177} (1986) 352;\\
P.~Wagemans, Phys.\ Lett.\ B {\bf 206} (1988) 241 and
{\it Aspects of N=4 supergravity}, Ph.D. Thesis,
Groningen University report RX-1299 (1990).

\bibitem{condflux}
  G.L. Cardoso, G. Curio, G. Dall'Agata, D. L\"ust,
  Fortsch. Phys. \textbf{52} (2004) 483
  [arXiv:hep-th/0310021];\\
  G.L. Cardoso, G. Curio, G. Dall'Agata, D. L\"ust,
  JHEP \textbf{0409} (2004) 059
  [arXiv:hep-th/0406118];\\
  L.~Gorlich, S.~Kachru, P.~K.~Tripathy and S.~P.~Trivedi,
  JHEP {\bf 0412} (2004) 074
  [arXiv:hep-th/0407130];\\
  K.~Choi, A.~Falkowski, H.~P.~Nilles, M.~Olechowski and S.~Pokorski,
  JHEP {\bf 0411} (2004) 076
  [arXiv:hep-th/0411066];\\
 P.~Manousselis, N.~Prezas and G.~Zoupanos,
  arXiv:hep-th/0511122.

\bibitem{ALT}
 G.~Dall'Agata and S.~Ferrara,
  Nucl.\ Phys.\ B {\bf 717}, 223 (2005)
  [arXiv:hep-th/0502066]; \\
  L.~Andrianopoli, M.~A.~Lledo and M.~Trigiante,
  JHEP {\bf 0505}, 051 (2005)
  [arXiv:hep-th/0502083]; \\
  G.~Dall'Agata, R.~D'Auria and S.~Ferrara,
  Phys.\ Lett.\ B {\bf 619}, 149 (2005)
  [arXiv:hep-th/0503122]; \\
  G.~Dall'Agata and N.~Prezas,
  JHEP {\bf 0510} (2005) 103
  [arXiv:hep-th/0509052].

\bibitem{threshold}
  L.~J.~Dixon, V.~Kaplunovsky and J.~Louis,
  Nucl.\ Phys.\ B {\bf 355}, 649 (1991);\\
  J.-P.~Derendinger, S.~Ferrara, C.~Kounnas and F.~Zwirner,
  Nucl.\ Phys.\ B {\bf 372}, 145 (1992) and
  Phys.\ Lett.\ B {\bf 271}, 307 (1991).

\bibitem{Mariosthreshold}
  E.~Kiritsis, C.~Kounnas, P.~M.~Petropoulos and J.~Rizos,
  Phys.\ Lett.\ B {\bf 385} (1996) 87
  [arXiv:hep-th/9606087] and
  Nucl.\ Phys.\ B {\bf 540} (1999) 87
  [arXiv:hep-th/9807067].

\bibitem{VY}
  G.~Veneziano and S.~Yankielowicz,
  Phys.\ Lett.\ B {\bf 113} (1982) 231.

\bibitem{Konishi}
  K.~Konishi,
  Phys.\ Lett.\ B {\bf 135} (1984) 439.

\bibitem{VZ}
  G. Villadoro and F. Zwirner,  hep-th/0508167.

\bibitem{BDFS}
F.~Buccella, J.-P.~Derendinger, S.~Ferrara and C.~A.~Savoy,
  Phys.\ Lett.\ B {\bf 115} (1982) 375; \\
 C.~Procesi and G.~W.~Schwarz,
  Phys.\ Lett.\ B {\bf 161}, 117 (1985).

\bibitem{DZ}
J.-P. Derendinger and A. Zaugg, in preparation.

\bibitem{noscale}
  E.~Cremmer, S.~Ferrara, C.~Kounnas and D.~V.~Nanopoulos,
  Phys.\ Lett.\ B {\bf 133} (1983) 61; \\
  J.~R.~Ellis, A.~B.~Lahanas, D.~V.~Nanopoulos and K.~Tamvakis,
  Phys.\ Lett.\ B {\bf 134} (1984) 429;\\
  J.~R.~Ellis, C.~Kounnas and D.~V.~Nanopoulos,
  Nucl.\ Phys.\ B {\bf 247} (1984) 373;\\
  A.~B.~Lahanas and D.~V.~Nanopoulos,
  Phys.\ Rept.\  {\bf 145} (1987) 1.

\end{thebibliography}
\end{document}